\documentclass[prd,superscriptsize,onecolumn,superscriptaddress,nofootinbib,notitlepage]{revtex4-1}
\usepackage[utf8]{inputenc} 

\usepackage{graphicx,xcolor,overpic,mathtools}
\usepackage{amsthm,amsmath,amssymb}
\usepackage[colorlinks]{hyperref}
\definecolor{coolblack}{rgb}{0.0, 0.18, 0.39}
\hypersetup{
    colorlinks = true,
    citecolor = {coolblack},  
    urlcolor = {blue}
   }
\PassOptionsToPackage{normalem}{ulem}
\usepackage{ulem} 
\usepackage{sidenotes}
\usepackage{bm}
\usepackage{url}
\usepackage{physics}
\usepackage[makeroom]{cancel}
\usepackage{tensor}
\usepackage{mathrsfs}
\usepackage{slashed}
\usepackage{tikz}
\usepackage[mode=buildnew]{standalone}
\usepackage{subcaption}

\newcommand{\comment}[1]{}

\NewDocumentCommand{\evat}{sO{\bigg}mm}{%
  \IfBooleanTF{#1}
   {\mleft. #3 \mright|_{#4}}
   {#3#2|_{#4}}%
}
\usepackage{enumerate}
\usepackage{cleveref}

\usepackage{stackengine}
\stackMath

\DeclareMathOperator{\Mev}{MeV}
\DeclareMathOperator{\fm}{fm}

\newcommand{\nn}{\nonumber\\ }
\newcommand{\beq}{\begin{eqnarray}}
\newcommand{\eeq}{\end{eqnarray}}

\begin{document}
\title[
]{
Nuclear mass radius and pressure in the Skyrme model}

\author{Alberto García Martín-Caro}
\email{alberto.garciam@ehu.eus}
\affiliation{Department of Physics, University of the Basque Country UPV/EHU, Bilbao, Spain}
\affiliation{Departamento de F\'isica de Part\'iculas, Universidad de Santiago de Compostela and Instituto
Galego de F\'isica de Altas Enerxias (IGFAE) E-15782 Santiago de Compostela, Spain}
\author{Miguel Huidobro}
\email{miguel.huidobro.garcia@usc.es}
\affiliation{Departamento de F\'isica de Part\'iculas, Universidad de Santiago de Compostela and Instituto
Galego de F\'isica de Altas Enerxias (IGFAE) E-15782 Santiago de Compostela, Spain}
 \author{Yoshitaka Hatta}
 \email{yhatta@bnl.gov}
\affiliation{Physics Department, Brookhaven National Laboratory, Upton, NY 11973, USA}
\affiliation{RIKEN BNL Research Center, Brookhaven National Laboratory, Upton, NY 11973, USA}

\begin{abstract} 
We compute the mass radius, scalar radius, tensor radius,   baryon number radius and mechanical radius of  nuclei with baryon number $B=1,2,3,4,5,6,7,8,32,108$ in the Skyrme model.  The relations between these radii and the nuclear gravitational form factors are  investigated. 
We also compute the  `pressure'  distribution  and find that it is negative in the core region for all the nuclei with $B>1$. This suggests that the way  mechanical stability  is achieved in nuclei is qualitatively different than in the nucleon. 

\end{abstract}

\maketitle

\section{Introduction}

Traditionally, since 1950s, the size of the proton has been determined through the measurement of the electromagnetic form factors in lepton-proton  elastic scattering \cite{Hofstadter:1956qs}. The {\it charge} radius thus extracted, about 0.8 fm, helps us to draw an intuitive picture of how quarks are distributed inside the proton. Its precise value is fundamentally important in QCD, QED and atomic spectroscopy. The same method can be used to define and determine the radius of atomic nuclei \cite{Hofstadter:1956qs,Frosch:1968zz,DeVries:1987atn,Sick:2008zza}. After years of controversy known as the proton radius puzzle  \cite{Carlson:2015jba,Gao:2021sml}, the most recent measurements have finally found consistency between the results from  electron scattering  and  muonic hydrogen (muonic atom) spectroscopy  
 \cite{Pohl:2010zza,Xiong:2019umf,Krauth:2021foz}.  

However, the charge radius is by no means the unique characterization of the size of hadrons and nuclei. For neutral hadrons, it may not be literally interpreted as a measure of physical size,  as exemplified by the negative mean square radius of the neutron 
\cite{Kopecky:1995zz}. Even for  charged hadrons, the charge radius does not fully reflect their internal  structure.  The virtual photon exchanged in electron scattering only sees  quarks and does not probe gluons which are essential constituents of the target. It then appears reasonable and  complementary to characterize the size of hadrons and nuclei in terms of the distribution of energy, angular momentum, etc.,   carried by quarks and gluons. This can be done by defining a radius  through the gravitational form factors (GFFs), { the hadronic form factors associated with the QCD energy-momentum tensor $T^{\mu\nu}$ } \cite{Kobzarev:1962wt,Pagels:1966zza}. 

Since $T^{\mu\nu}$ has many components,  one can define various radii with different physical interpretations \cite{Goeke:2007fp,Cebulla:2007ei}. For example, the {\it mass} radius measures the distribution of the energy density $T^{00}({\bm x})$. Similarly, the {\it scalar} radius is associated with the QCD trace anomaly density $T^\mu_\mu({\bm x})$.   An obvious question, however, is whether they are measurable in experiments. In fact, it is only recently that there have been new ideas and attempts at extracting the gravitational form factors and the associated radii \cite{Kumano:2017lhr,Burkert:2018bqq,Hatta:2018ina,Mamo:2019mka,Kharzeev:2021qkd,Sun:2021gmi,Duran:2022xag,Wang:2022vhr,Guo:2023pqw,Wang:2023uek,GlueX:2023pev,Hatta:2023fqc}. While at the moment the precision of such extractions falls behind that of the electromagnetic form factors, this is an encouraging development worth pursuing in the future, especially towards the era of the  Electron-Ion Collider (EIC) \cite{AbdulKhalek:2021gbh}.  

Just like the charge radius, the discussion  of the GFFs and associated radii can be extended to atomic  nuclei. 
In a previous paper \cite{GarciaMartin-Caro:2023klo}, we have computed, for the first time, the GFFs of various nuclei within the Skyrme model \cite{Skyrme:1961vq}. In this model, a nucleus is realized as a classical field configuration (`Skyrmion') with a definite baryon number $B$. The calculation of the GFFs is complicated by the fact that the classical solutions are not spherically symmetric for $B>1$, but we have successfully extracted one of the GFFs, the $D(t)$ form factor,  for a variety of nuclei.  We find that  the so-called D-term $D(t=0)\sim B^{1.8}$ grows rapidly with increasing $B$. (Compare with the earlier works \cite{Polyakov:2002yz,Liuti:2005qj,Guzey:2005ba} and a recent work \cite{He:2023ogg}).) As the D-term is commonly associated with the internal  radial force, or `pressure', it affects the size of the system.  The goal of this paper is to systematically compute various definitions of radii  for various nuclei, and study their connection to the D-term. 

\section{Radius zoo}

In this section, we introduce a number of radii for the proton. 
Let us start with the familiar electromagnetic form factors which are the matrix element of the electromagnetic current $J_{em}^\mu=\sum_q e_q \bar{q}\gamma^\mu q$ ($e_u=\frac{2}{3}, e_d=-\frac{1}{3}$, etc.)
\beq
\langle p'|J^{\mu}_{em}|p\rangle = \bar{u}(p')\left[\gamma^\mu F_1(t) + \frac{i\sigma^{\mu\rho}\Delta_\rho}{2M}F_2(t)\right]u(p),
\eeq 
where { $\sigma^{\mu\nu}=\frac{i}{2}[\gamma^\mu,\gamma^\nu]$ and $u(p)$ is the proton spinor. } $\Delta^\mu = p'^\mu -p^\mu$, $t=\Delta^2$ and $M$ is the proton mass. In  the Breit frame where $\Delta^0=0$ and $t=-{\bm \Delta}^2$, 
\beq
\langle p'|J^{0}_{em}|p\rangle = 2M \left(F_1(t)+\frac{t}{4M^2}F_2(t)\right)\equiv 2MG_E(t).
\eeq
Introducing the charge density 
\beq
\rho_c({\bm x})= \int \frac{d{\bm \Delta}}{(2\pi)^3}e^{i{\bm x}\cdot {\bm \Delta}} G_E(t),
\eeq
 one defines the  charge radius as  
\beq
\left.\langle r^2\rangle_c =\frac{\int d{\bm x} x^2\rho_c({\bm x})}{\int d{\bm x}\rho_c({\bm x})}=   \frac{6}{G_E(0)}\frac{dG_E(t)}{dt}\right|_{t=0}.
\eeq
Similarly, one can introduce form factors for the baryon number current $J_B^\mu =\frac{1}{3}\sum_q\bar{q}\gamma^\mu q$ 
and the corresponding radius 
\beq
\langle p'|J^0_{B}|p\rangle = 2MG_{B}(t), \qquad \rho_{B}({\bm x}) = \int \frac{d{\bm \Delta}}{(2\pi)^3}e^{i{\bm x}\cdot {\bm \Delta}} G_{B}(t), \qquad  \left.\langle r^2\rangle_{B} =\frac{\int d{\bm x} x^2\rho_{B}({\bm x})}{\int d{\bm x}\rho_B({\bm x})}= \frac{6}{G_B(0)}\frac{dG_{B}(t)}{dt}\right|_{t=0}.\label{rhob1}
\eeq
For the proton, $\int d{\bm x} \rho_{c,B}({\bm x})=1$. 
One may also introduce radii associated with the isospin and axial vector currents.

Next, the gravitational form factors of the nucleon { are defined as} 
\beq
\langle p'|T^{\mu\nu}|p\rangle =\bar{u}(p')\left[ \gamma^{(\mu} P^{\nu)} A(t) + \frac{iP^{(\mu}\sigma^{\nu)\rho}\Delta_\rho}{2M}B(t)+\frac{D(t)}{4M}(\Delta^\mu \Delta^\nu -g^{\mu\nu}\Delta^2)\right]u(p),
\eeq
where, $g^{\mu\nu}={\rm diag}(1,-1,-1,-1)$,   $P^\mu = \frac{p^\mu+p'^\mu}{2}$ and the brackets $(\mu\nu)$ denote the symmetrization of indices. Again in the Breit frame, we define the spatial distribution 
\beq
T^{\mu\nu}({\bm x}) \equiv \int \frac{d{\bm \Delta}}{(2\pi)^3} e^{i{\bm x}\cdot {\bm \Delta}}\frac{\langle p'|T^{\mu\nu}|p\rangle}{\bar{u}(p')u(p)}. \label{energy}
\eeq
Since $\int d{\bm x}T^{00}({\bm x})=M$, $T^{00}({\bm x})$ can be interpreted as the mass density. This leads to the definition of the mass radius  \cite{Goeke:2007fp,Cebulla:2007ei}
\beq
\left.\langle r^2\rangle_m = \frac{\int d{\bm x}\, x^2 T^{00}({\bm x})}{\int d{\bm x} T^{00}({\bm x})} = 6\frac{dA(t)}{dt}\right|_{t=0}- \frac{3D(0)}{2M^2},
\label{mass}
\eeq
where we used  $B(0)=0$ to get the right hand side. 
We also consider the scalar radius associated with the trace of the energy momentum tensor $T^\mu_\mu$ 
\beq
\left.\langle r^2\rangle_s = \frac{\int d{\bm x}\, x^2 T^\mu_\mu({\bm x})}{\int d{\bm x} T^\mu_\mu({\bm x})}= 6\frac{dA(t)}{dt}\right|_{t=0}- \frac{9D(0)}{2M^2} ,
\label{scalar}
\eeq
and the tensor radius 
\beq 
\left.\langle r^2\rangle_t \equiv \frac{\int d{\bm x}\, x^2 \left(T^{00}({\bm x})+\frac{1}{2}T_{ii}({\bm x})\right)}{\int d{\bm x} \left(T^{00}+\frac{1}{2}T_{ii}\right)} =6\frac{dA(t)}{dt}\right|_{t=0}. \label{ten}
\eeq
($T_{ii}$ is a short for $\sum_{i=1}^3T_{ii}$, so that $T^\mu_\mu=T^{00}-T_{ii}$.) The linear combination in (\ref{ten}) is understood as follows (see, also, Section 5 of  \cite{Fujita:2022jus}). A scalar (spin-0) exchange with momentum $\Delta^\mu$ excites a mode   $T^{\mu\nu}\sim g^{\mu\nu}-\frac{\Delta^\mu \Delta^\nu}{\Delta^2}$  in the energy momentum tensor. In the Breit frame $\Delta^\mu=(0,{\bm \Delta})$, $T^{00}\sim 1$, and $T^{ij} \sim -\delta^{ij}+\frac{\Delta^i \Delta^j}{{\bm \Delta}^2}$. Therefore, by forming the linear combination $T^{00}+\frac{1}{2}T_{ii}$, one can eliminate the spin-0 component.

A few comments are in order. First,  the denominators in (\ref{mass})-(\ref{ten}) are equal to $M$ because $\int d{\bm x}T_{ij}({\bm x})=0$ as a consequence of  the conservation law $\partial_iT_{ij}=0$, { namely, $0=\int d{\bm x} x_i\partial_kT_{kj}=-\int d{\bm x}T_{ij}$. Such an identity can be interpreted also as a consequence of Derrick's scaling theorem for soliton configurations \cite{Manton:2008ca}}. Second, the three radii are not independent of each other. They differ by the so-called D-term $D(t=0)$   
\beq
\langle r^2\rangle_m = \frac{2}{3}\langle r^2\rangle_t +\frac{1}{3}\langle r^2\rangle_s, \qquad \langle r^2\rangle_s = \langle r^2\rangle_m -\frac{3D(0)}{M^2}. \label{diff}
\eeq
If  $D(0)$ is negative for the proton, as is generally believed, then we have the ordering $\langle r^2\rangle_t < \langle r^2\rangle_m < \langle r^2\rangle_s$. Finally, our definition of the tensor radius (\ref{ten}) differs from that in Ref.~\cite{Ji:2021mtz} which reads 
\beq  
\left.\langle r^2\rangle'_t \equiv \frac{\int d{\bm x}\, x^2 \left(T^{00}({\bm x})+\frac{1}{3}T_{ii}({\bm x})\right)}{M} =6\frac{dA(t)}{dt}\right|_{t=0}-\frac{D(0)}{2M^2}. \label{ji}
\eeq 
The difference is due to the meaning of the word `tensor'. (\ref{ji}) is based on the  decomposition of the energy momentum tensor into the traceless and trace parts  
\beq
T^{\mu\nu}=\left(T^{\mu\nu}-\frac{g^{\mu\nu}}{4}T^\alpha_\alpha\right) + \frac{g^{\mu\nu}}{4}T^\alpha_\alpha,
\qquad \to \qquad 
T^{00}=\frac{3}{4} \left(T^{00}+\frac{1}{3}T_{ii}\right)+\frac{1}{4}(T^{00}-T_{ii}). \label{traceless}
\eeq 
The first term is spin-2 (tensor) and the second is spin-0 (scalar), where `spin' refers to the Lorentz spin. On the other hand,  hadrons are classified according to the spin quantum numbers  $J=0,\frac{1}{2},1,\frac{3}{2},2,\cdots$ of the SU(2) group.  The traceless part in (\ref{traceless}) does not exactly correspond to the irreducible $J=2$ component. The latter has to be  transverse-traceless,\footnote{ { A transverse-traceless tensor $X^{\mu\nu}$ satisfies $\partial_\mu X^{\mu\nu}=X^\mu_\mu=0$. When $X$ is symmetric, these conditions reduce its independent components  to five,  which correspond to the physical degrees of freedom of a spin-2 particle.}} not just traceless, and the relevant  decomposition is  
\beq
T^{\mu\nu}= \left\{ T^{\mu\nu}-\frac{1}{3}\left(g^{\mu\nu} -\frac{\partial^\mu \partial^\nu}{\partial^2}\right)T^\alpha_\alpha\right\} + \frac{1}{3}\left(g^{\mu\nu} -\frac{\partial^\mu \partial^\nu}{\partial^2}\right)T^\alpha_\alpha,
\eeq
{ where $\frac{1}{\partial^2}T^\alpha_\alpha(x)=\int dy\frac{-i}{4\pi^2(x-y)^2}T^\alpha_\alpha(y)$. (In momentum space $x^\mu \to \Delta^\mu$ with $\Delta^0=0$, we can simply write $\frac{1}{\partial^2}\to \frac{1}{{\bm \Delta}^2}$.) }
In our definition (\ref{ten}), the tensor radius is exclusively linked to $J^{PC}=2^{++}$ tensor hadrons (in particular, glueballs \cite{Mamo:2022eui,Fujita:2022jus}) which saturate the $A(t)$-form factor. On the other hand, the $D(t)$-form factor, hence also (\ref{ji}), receives contributions from both $2^{++}$ and $0^{++}$ hadrons. 

Finally, we also consider the so-called mechanical radius defined solely by the $D(t)$ form factor  \cite{Polyakov:2018zvc} 
\beq
\langle r^2\rangle_{mech} = \frac{\int d{\bm x} x^2 \frac{x_ix_j}{x^2}T_{ij}({\bm x})}{\int d{\bm x} \frac{x_ix_j}{x^2}T_{ij}({\bm x})} = \frac{6D(0)}{\int_{-\infty}^0 dt D(t)}.   \label{mech}
\eeq
The projection $\frac{x_ix_j}{x^2}T_{ij}$ may be interpreted as the momentum flux in the radial direction.  
Thus the mechanical radius measures the mean square radius of the distribution of `radial force'.   

\section{Nuclear radii in the Skyrme model}

We now lay out our strategy to compute the various radii introduced in the previous section for atomic nuclei specifically in the context of the Skyrme model. In this model, a nucleus is realized as an SU(2)-valued classical field configuration (`Skyrmion') $U({\bm x})$ which satisfies the equation of motion and  carries an integer baryon number $B=1,2,3,\cdots$. The solution is then quantized via the method of collective coordinate quantization. Solutions up to $B=108$ have been known for some time, and their quantum properties, such as the excitation spectrum, have been studied  \cite{Feist:2012ps,BjarkeGudnason:2018bju}.   Very recently, new solutions up to $B=256$ have been constructed \cite{GarciaMartin-Caro:2023pjm} using the new Skyrmions3D package \cite{Halcrow_Skyrmions.jl}.  We have numerical  solutions with $B=1,2,3,4,5,6,7,8,32,108$ available at hand. From them, it is easy to compute the energy momentum tensor density $T^{\mu\nu}({\bm x})$, to be identified with (\ref{energy}), as well as   the baryon number density (\ref{rhob1})
\begin{equation}
    \rho_B({\bm x}) = \frac{\epsilon^{0\nu\rho\sigma}}{24\pi^2}\Tr\{(U^\dagger \partial_{\nu}U)(  U^\dagger\partial_{\rho} U)( U^\dagger \partial_{\sigma}U)\}, \qquad \int d{\bm x} \rho_B({\bm x})=B, \label{rhob}
\end{equation}
We can then  evaluate  $\langle r^2\rangle_{B}$ and $\langle r^2\rangle_{m,s,t,mech}$  directly in the coordinate space. 
 The calculation of the charge radius or the isospin radius $\langle r^2\rangle_{c,I}$ is more involved because it requires the spin-isospin quantization procedure  specific to individual nuclei. Therefore, results for only the lightest Skyrmion solutions, i.e., $B=2$ (deuteron) and $B=3$ (triton and helium-3) can be found in the literature \cite{Braaten:1988bn,Carson:1991fu}.  (See also the recent discussion \cite{GarciaMartin-Caro:2023pjm} on an observable related to these densities, the neutron skin thickness, and its computation for larger Skyrmions.) In these works, it was observed that the Skyrme model predictions undershoot the experimental data. Namely, the observed charge radii of the proton and the $B=2,3$ nuclei differ by a factor of 2 or more \cite{DeVries:1987atn}, whereas the model tends to predict a smooth, gradual increase with $B$. Also, the binding energies of light nuclei come out to be too large. The resolution of these issues requires various modifications of the model, see e.g., \cite{Gillard:2015eia,Gudnason:2018jia}.  
Since the calculation of $\langle r^2\rangle_m$, etc.  for $B>1$ nuclei is the first study of this kind, and moreover it does not require the quantization procedure, we do not consider such modifications in this paper.  Instead, we shall treat the proton charge radius $\sqrt{\langle r^2\rangle_c}=0.84$ fm as an input to fix the model parameters.

Before presenting numerical results, we need to check  whether the relations   between various radii and the D-term discussed in the previous section remain the same for nuclei.  This is nontrivial because the Skyrmions are not spherically symmetric for $B>1$. Related to this,  nuclei come with various spins $J=0,\frac{1}{2},1,\cdots$. For $J=0$, the $A,D$ form factors are defined similarly, but for $J>\frac{1}{2}$, there are many more form factors characterizing the nuclear deformation. Nevertheless, we have shown in \cite{GarciaMartin-Caro:2023klo} 
 that, even for deformed nuclei with nonspherical $T_{ij}$, one can compute the `monopole' D-term  via the same formula 
\begin{eqnarray}
  \frac{D(t)}{M}&=&6\int d{\bm x} \left(x^ix^j-\frac{1}{3}\delta^{ij} x^2\right)\frac{j_2(\sqrt{-t}x)}{t x^2}T_{ij}({\bm x}) , \label{main}
\end{eqnarray}
\beq
\frac{D(0)}{M} = -\frac{2}{5}\int d{\bm x} \left(x_i x_j-\frac{1}{3}\delta_{ij}x^2\right)T_{ij}({\bm x}) ,
\eeq
as for the  $B=1$ solution. { ($j_2$ is the spherical Bessel function).} Now let us combine this observation with the relation
\beq
\int d{\bm x} x_i x_j T_{ij}({\bm x}) = -\frac{1}{2}\int d{\bm x}x^2 T_{ii}({\bm x}),
\eeq
which can be easily proven by using the conservation law $0=\int d{\bm x}x^2 x_i\partial_j T_{ij}$ and partial integration. We find a formula valid for any $B$,  
\beq
\frac{D(0)}{M} = -\frac{2}{3} \int d{\bm x} x_i x_j T_{ij}({\bm x})=\frac{1}{3}\int d{\bm x} x^2T_{ii}({\bm x}).  \label{new}
\eeq
Moreover, from (\ref{main}), we find 
\beq
\int_{-\infty}^0 dt \frac{D(t)}{M} = -4\int d{\bm x} \left(\frac{x^ix^j}{x^2}-\frac{1}{3}\delta^{ij} \right)T_{ij}({\bm x}) = -4\int d{\bm x} \frac{x^ix^j}{x^2}T_{ij}({\bm x}). \label{new2}
\eeq 
We see that (\ref{new}) and (\ref{new2}) are compatible with  (\ref{mass})-(\ref{ten}) and (\ref{mech}).

In addition to radii, we also consider the `pressure'
distribution\footnote{ { For nuclei with $J=1$ or larger, in general `pressure' can be anisotropic due to nuclear deformation. Here we only consider the isotropic component defined by the monopole D-term (\ref{main}) mentioned above.}}     \cite{Polyakov:2002yz,Goeke:2007fp,Cebulla:2007ei}
\begin{equation}
    p(r)=\frac{1}{6M}\int \frac{d{\bm \Delta}}{(2\pi)^3}e^{i{\bm \Delta}\cdot {\bm r}}tD(t) =  \frac{1}{24\pi^2M} \int^0_{-\infty} dt \frac{\sin \sqrt{-t} \, r}{r} tD(t) . \label{pres}
\end{equation}
While it is straightforward to evaluate this using our result for $D(t)$ \cite{GarciaMartin-Caro:2023klo}, we can further simplify it by rewriting  (\ref{main}) as  
\beq
\frac{tD(t)}{M} &=& -6 \int d{\bm x}\left(x_ix_j-\frac{\delta_{ij}}{3}x^2\right) \frac{1}{x}\frac{d}{dx}\left(\frac{j_1(\sqrt{-t} x)}{\sqrt{-t}x}\right)T_{ij}({\bm x}) \nn 
&=& -\frac{6}{\sqrt{-t}}\int d{\bm x} \left(x_i\partial_j\left(\frac{j_1(\sqrt{-t} x)}{x}\right)T_{ij}({\bm x}) -\frac{x}{3} \frac{d}{dx}\left(\frac{j_1(\sqrt{-t} x)}{x}\right) T_{ii}({\bm x}) \right) \nn 
&=& 2\int d{\bm x} j_0(\sqrt{-t}x)T_{ii}({\bm x}), \label{simpler}
\eeq
where we used the conservation law $\partial_jT_{ij}=0$ and the formulas $\frac{d}{dx}(j_1(x)/x)= -j_2(x)/x$, $\frac{d}{dx}(x^2j_1(x))= x^2j_0(x)$. 
Substituting  (\ref{simpler}) into (\ref{pres}) and using the orthogonality relation 
$\int_0^\infty dz z^2 j_0(rz)j_0(xz)= \frac{\pi}{2r^2}\delta(x-r)$, we find 
\beq
p(r)= \frac{1}{12\pi} \int d\Omega\,  T_{ii}(|{\bm x}|=r), \label{ni}
\eeq
where $d\Omega$ denotes solid angle integration. 
In the spherically symmetric case, we recover the familiar formula $p(r)=\frac{1}{3}T_{ii}(r)$.

\comment{
\begin{figure}
    \centering
    \includegraphics[scale=0.55]{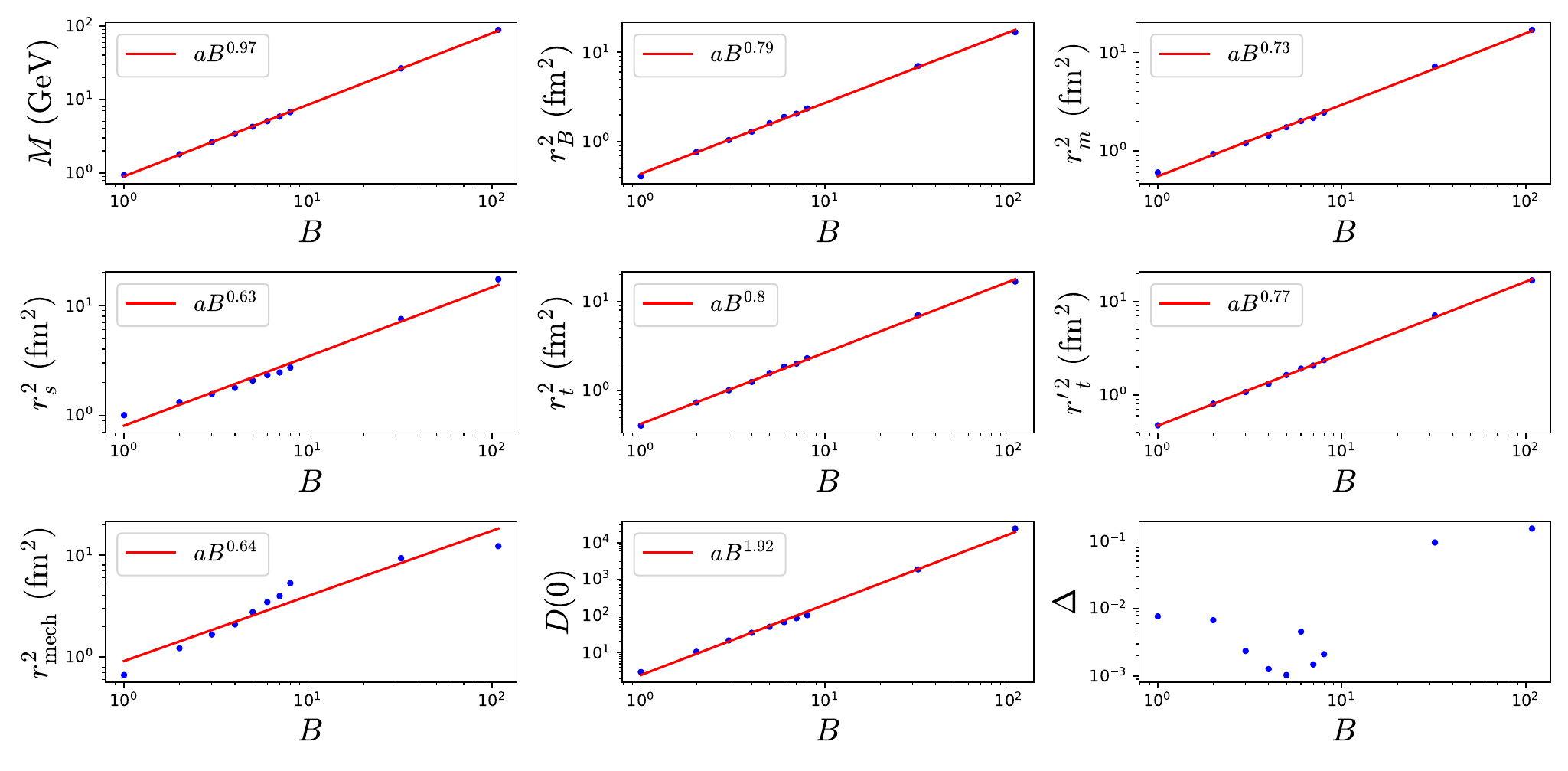}
    \caption{Dependence of various quantities on the  baryon number.  We also plot the mass $M$ and  $\Delta$ as defined in \eqref{deltaeq}.}
    \label{fig:radii_D}
\end{figure}
}
\begin{figure}
    \centering
    \includegraphics[scale=0.6]{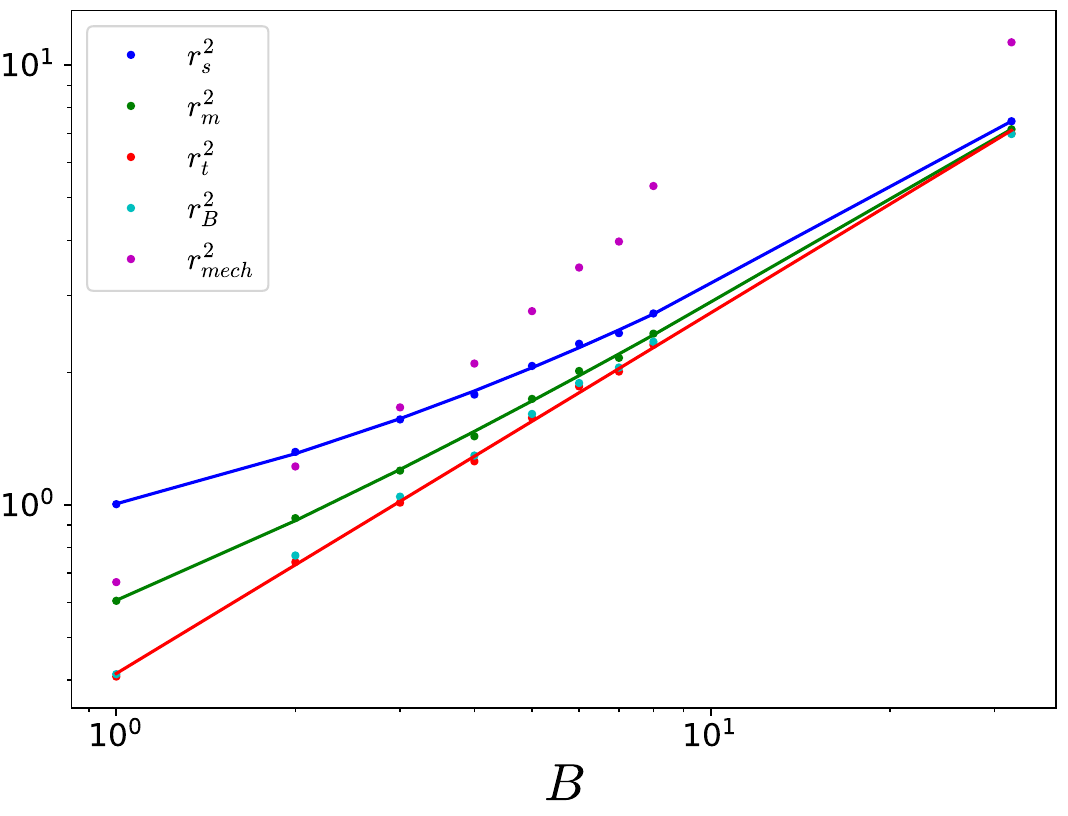}
    \caption{Values of $\langle r^2_{s,m,t,B}\rangle$ and their best fit curve, together with $\langle r^2_{\rm mech}\rangle$. Note that $\langle r^2_{t}\rangle$ and $\langle r^2_{B}\rangle$ coincide almost exactly, hence we have used the same fitting parameters for both radii.}
    \label{fig:rmstb}
\end{figure}
\begin{figure}
\centering    \includegraphics[scale=0.6]{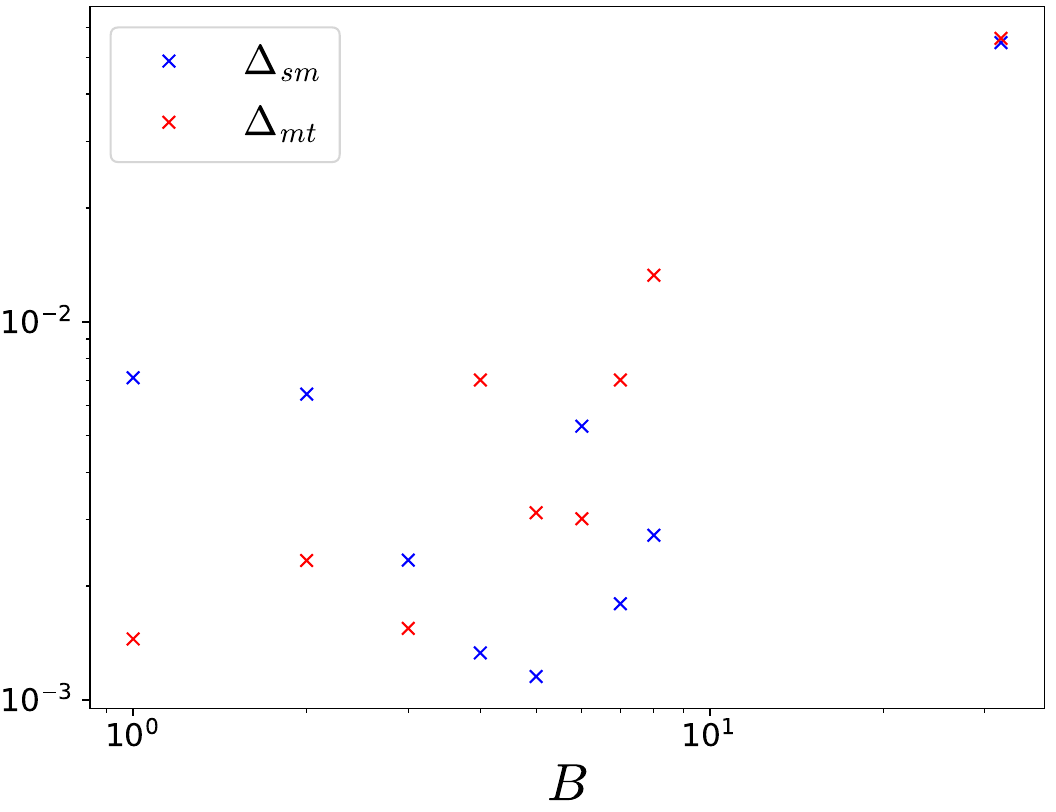}
    \caption{Residual error for the relations between $r_s,r_m$ and $r_t$.}
    \label{fig:enter-label}
\end{figure}

\section{Numerical results}

We use the original Skyrme model Lagrangian supplemented with the pion mass term.  Its precise form as well as our numerical approach is explained   in \cite{GarciaMartin-Caro:2023klo} to which we refer the reader.  In order to numerically confirm relations such as (\ref{diff}) which are sensitive to the {\it difference} of radii, we need more accurate solutions than those presented in \cite{GarciaMartin-Caro:2023klo}. This has been achieved by enlarging the 3D grid (up to $N^3=(151)^3$ points) and accelerating the energy-minimization process via a parallelization of the numerical algorithm using the \textit{OpenMP} C++ package.   Thanks to this improvement, the residual errors  { (cf. (\ref{diff})) } 
\begin{equation}
    \Delta_{sm}=\abs{\frac{\langle r^2\rangle_m-\langle r^2\rangle_s}{\frac{3D(0)}{M^2}}-1},\qquad \Delta_{mt}=\abs{\frac{\langle r^2\rangle_t-\langle r^2\rangle_m}{\frac{3D(0)}{2M^2}}-1},
    \label{deltaeq}
\end{equation}
is less than 1\% for the solutions $1\le B \le 8$, { with $\Delta\sim 5\%$ for the $B=32$} and $\Delta\lesssim 10\%$ for the $B=108$ solutions.  We have then fixed the  parameters of the model (see Eq.~(1) of \cite{GarciaMartin-Caro:2023klo}) as $f_\pi=117.23\Mev,$ $e=4.81$ and $m_\pi=138 \Mev$ such that the proton's mass $M=0.938$ GeV and  charge radius $\sqrt{\langle r^2\rangle_c}=0.84$ fm are reproduced. The sextic coupling discussed in \cite{GarciaMartin-Caro:2023klo} has been turned off. 
Below  we introduce a simpler notation  for the root mean square radii $\sqrt{\langle r^2\rangle_m}\equiv r_m$, etc. 
In Fig.~\ref{fig:rmstb}, we plot the radii $r_m$, $r_s$, $r_t$, $r_{mech}$ and $r_B$   as a function of $B$.\footnote{ { See \cite{Flambaum:2023yua} for an earlier calculation of $r_B$ with $B>1$ from approximate Skyrmion solutions.} } The actual values are complied in Table~\ref{tab:radii_data} together with the masses $M$ and the  values of $D(0)$ (different from \cite{GarciaMartin-Caro:2023klo} due to different parameters). We confirm the expected ordering $r_t<r_m<r_s$.  Curiously, $r_t$ and $r_B$ agree within less than 2\%. { (See \cite{Hagiwara:2024wqz} for an argument which might be related to this.) 
 The $B$-dependence is roughly consistent with the usual formula $r\sim B^{1/3}$, but some radii cannot be fitted well with a naive assumption  $r\sim B^a$ with a single exponent $a$.    Indeed, the relations (\ref{mass}) and (\ref{scalar}) imply that at least two exponents are involved $r_{m,s,t}\sim c_1 B^a+c_2 B^b$.} The difference between any two of them behaves as 
\beq
\frac{D(0)}{M^2} \propto B^{\beta-2},
\eeq
with  $\beta \approx 1.92$. (A smaller value  $\beta=1.7\sim 1.8$ was obtained in \cite{GarciaMartin-Caro:2023klo} with a different set of parameters.) Thus the differences go to zero as $B\to \infty$.
A similar behavior is expected for the model of \cite{Liuti:2005qj} which predicted $\beta\approx 1$.  In contrast,  the differences increase with $B$ in the models of \cite{Polyakov:2002yz,Guzey:2005ba} where   $\beta\approx 2.3$.

\begin{table}[t]
\centering
\begin{tabular}{cccccccccc}
\toprule
$B$ & $M$ (GeV) & $r_B$ & $r_m$ & $r_s$ & $r_t$ & $r'_t$ & $r_{\rm mech}$ & $D(0)$ & $p(0)$ (GeV/fm\textsuperscript{3}) \\
    \hline
    1 & 0.938 & 0.642 & 0.778 & 1.002 & 0.638 & 0.688 &  0.817 & -2.98 & 0.3465 \\
    2 & 1.795 & 0.876 & 0.966 & 1.149 & 0.861 & 0.898 & 1.106 & -10.61 & -0.0362 \\
    3 & 2.616 & 1.022 & 1.094 & 1.251 & 1.006 & 1.036 &  1.291 & -21.62 & -0.0203 \\
    4 & 3.408 & 1.138 & 1.197 & 1.335 & 1.121 & 1.147 &  1.448 & -34.78 & -0.0173 \\
    5 & 4.254 & 1.269 & 1.32 & 1.439 & 1.256 & 1.278 & 1.661 & -50.92 & -0.0185 \\
    6 & 5.064 & 1.376 & 1.42 & 1.525 & 1.364 & 1.383 & 1.862 & -68.25 & -0.0196 \\
    7 & 5.845 & 1.433 & 1.47 & 1.568 & 1.418 & 1.436 & 1.993 & -87.24 & -0.0173 \\
    8 & 6.698 & 1.533 & 1.566 & 1.651 & 1.521 & 1.536 & 2.305 & -105.31 & -0.0183 \\ 
    32 & 26.419 & 2.641 & 2.674 & 2.731 & 2.645 & 2.655 & 3.358 & -1745.44 & 0.0000 \\
    108 & 88.311 & 4.077 & 4.114 & 4.165 & 4.088 & 4.097 & 3.499 & -24151.55 & -0.0171\\
    \hline
\end{tabular}
\caption{ Root mean square radii of Skyrmions  in units of $\fm$. The proton charge radius $r_c=0.84$ fm is used as an input.}
\label{tab:radii_data}
\end{table}

\begin{figure}
    \centering
    \includegraphics[scale=0.52]{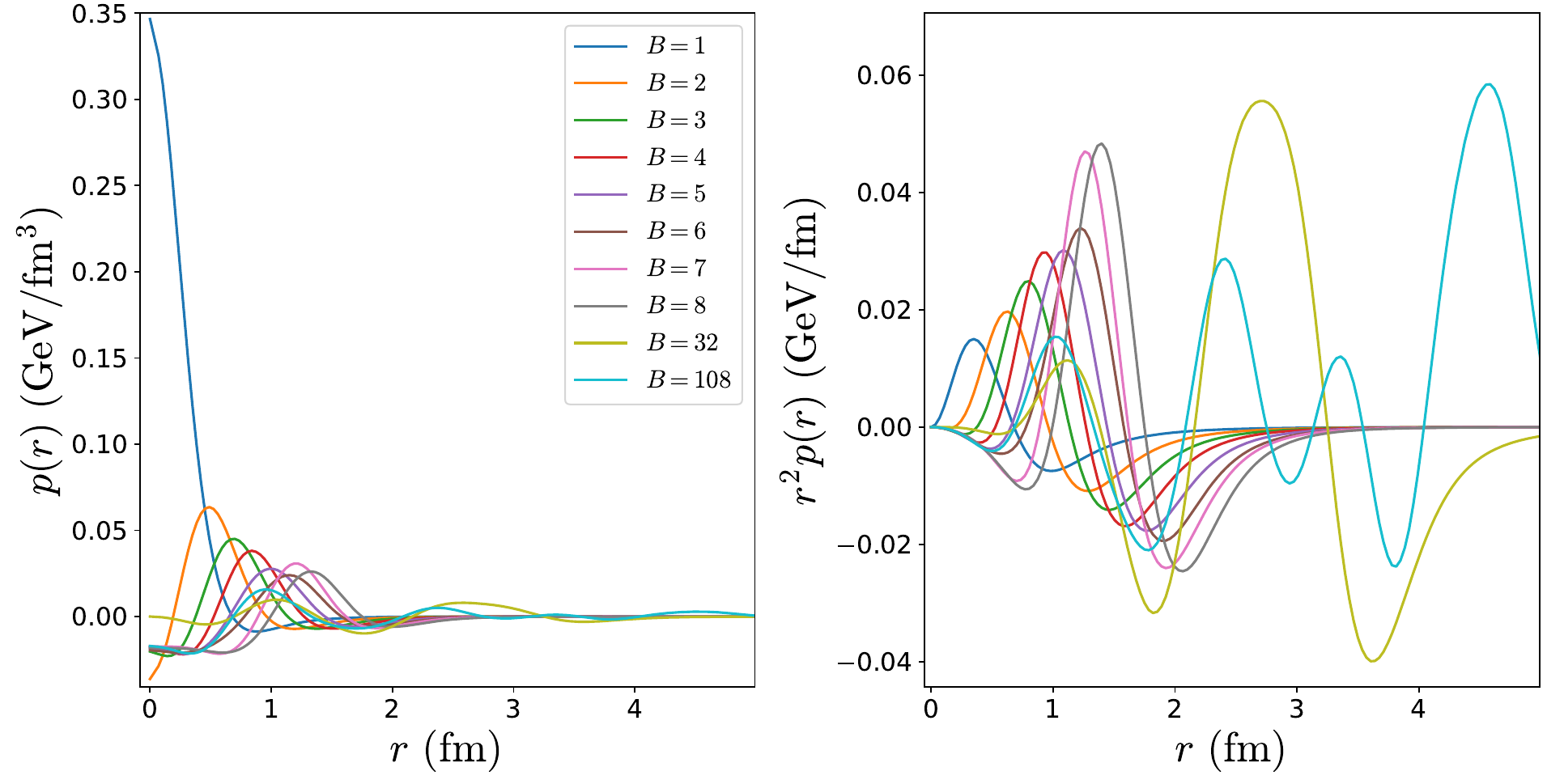}
    \caption{Effective radial pressure density for Skyrmions with $B=1 - 8$ and $B = 32$. }
    \label{fig:p_r}
\end{figure}

In Fig.~\ref{fig:p_r}, we plot the `pressure' distribution $p(r)$ (left) and $r^2p(r)$ (right).  We numerically checked consistency between the two definitions (\ref{pres}) and (\ref{ni}). The latter computation is tricky because it involves an integral over the solid angle, whereas our solutions are obtained in a 3D grid of Cartesian coordinates. Nevertheless, we have achieved a good numerical accuracy in performing this integral.   
Remarkably, we find that $p(0)$ is negative for all the solutions with $B>1$ except for $B=32$ where $p(0)$ is consistent with zero up to ${\cal O}(10^{-6})$.   {  Actually, the result $p(0)<0$ has been recently found   in \cite{Freese:2022yur} for the deuteron ($B=2$)\footnote{Ref.~\cite{Freese:2022yur} computed $p(r)$ for the three deuteron spin states and found $p(0)<0$ in one case. In our calculation, the different spin states are averaged over. } } and also in \cite{He:2023ogg}   for the helium-4 ($B=4$) with one  choice of the nuclear wavefunction. This is at first sight surprising and even unlikely because, from (\ref{pres}), 
\beq
p(0) = \frac{-1}{24\pi^2M}\int_{-\infty}^0 dt (-t)^{\frac{3}{2}}D(t).
\eeq
If $D(t)$ is negative definite, clearly $p(0)$ is positive. However, as already observed in \cite{GarciaMartin-Caro:2023klo}  (see also \cite{Freese:2022yur}), $D(t)$ turns positive in the  large-$|t|$ region for the $B>1$ solutions, and due to the weight factor $(-t)^{\frac{3}{2}}$, this region dominates the integral in practice.  A more intuitive way to understand this phenomenon is to notice that the solutions with $B>1$ have a `hollow' in the core region where the energy density is lower than in the surrounding region . This results in an inward flux near the origin which may be interpreted as negative pressure. 
For light nuclei $2\le B\le 8$, $p(r)$ then turns positive in the intermediate $r$-region and becomes negative again at large-$r$. Pressure per solid angle $r^2p(r)$ has a stronger peak at an increasingly large value of $r$ as $B$ is increased. In the case of large nuclei $B=32,108$,  $p(r)$ shows a striking oscillating pattern, similar to the oscillation of $D(t)$ found in  \cite{GarciaMartin-Caro:2023klo}. This is probably due to the fact that, in the present model, these nuclei are realized as  an `$\alpha$-cluster',  an organized spatial arrangement of $B=4$  solutions, see Fig.~1 of \cite{GarciaMartin-Caro:2023klo}.\footnote{ This may also explain the  exceptional result $p(0)\approx 0$ for the $B=32$ solution because the point $r=0$ is not populated by an $\alpha$-particle in this solution. On the other hand, in the $B=108$ solution, an $\alpha$-particle is positioned at  $r=0$.  
}  The energy density also oscillates as a function of $r$.

A common interpretation of the $p(r)$ curve of the nucleon (see the $B=1$ curve in Fig.~\ref{fig:p_r}) is that the `confining' force $p(r)<0$ in the outer region is counteracted by the `repulsive core' $p(r)>0$ near the origin \cite{Goeke:2007fp,Cebulla:2007ei,Burkert:2018bqq,Lorce:2018egm}. This provides an appealing scenario of how the nucleon may be  stabilized. However, such an interpretation does not hold verbatim for $B>1$ nuclei especially when $B\gg 1$,     suggesting that  the mechanical stability argument for nuclei is more nontrivial. { (See also \cite{Freese:2022yur} for a related discussion in the special case $B=2$.) }
One should also keep in mind that the `pressure' (\ref{pres}) defined by the $D(t)$-form factor is not genuine thermodynamic pressure. The existence of negative regions $p<0$  is a  reminder of this caveat.    

\begin{figure}
    \centering
    \includegraphics[scale=0.5]{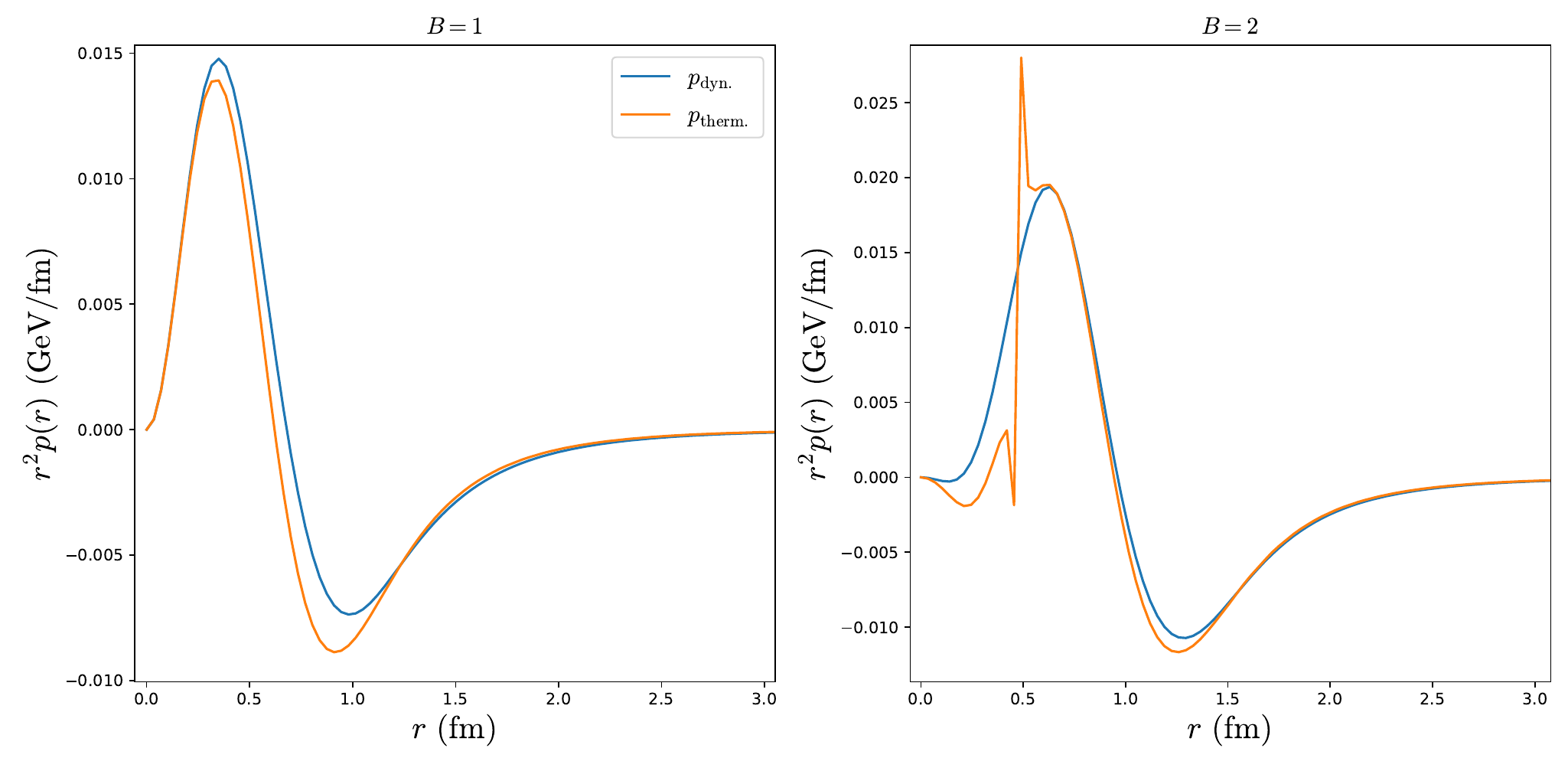}
    \caption{Pressure defined by (\ref{th_pres}) (orange) and by (\ref{pres}) (blue) for the $B=1$ (left) and $B=2$ (right) solutions. In the right panel, numerical accuracy is lost for $r\lesssim 0.6 \fm$ due to the error induced by the non-exact cancelation of vanishing radial derivatives in \eqref{th_pres}. Still, appreciable differences between both pressure definitions appear for larger values of the radius.}
    \label{fig:Pressures}
\end{figure}

In order to see the nature of  (\ref{pres}) and its limitation  from a different perspective, let us naively assume the standard thermodynamical relation between the pressure, baryon number (\ref{rhob}) and energy $\varepsilon=T^0_0$ densities  of a zero-temperature, barotropic fluid \cite{danielewicz2002determination}
\begin{equation}
p(\rho_B)=\rho_B^2\pdv{(\varepsilon/\rho_B)}{\rho_B}.
\label{p_def}
\end{equation}
 Assuming spherical symmetry, one could write an alternative expression for $p(r)$ 
\begin{equation}
    p(r)=\frac{\rho_B(r)}{\rho_B'(r)}\varepsilon'(r)-\varepsilon(r).
    \label{th_pres}
\end{equation}
This is plotted in Fig.~\ref{fig:Pressures} (left)   together with (\ref{pres}) for $B=1$. It can be seen that the two definitions of pressure  do not  coincide. This is not surprising, as the energy and baryon number densities have been computed independently of each other, and {\it a priori}  there is no  reason why  \eqref{th_pres} should coincide with (\ref{pres}). (We are however surprised that  the two curves are rather close.) Note also that (\ref{p_def}) asserts that the on-shell Skyrmion configurations satisfy the  barotropic equation of state $p=p(\rho_B)$, which   has been argued to not be the case for Skyrmion solutions in the context of the  equation of state of neutron stars \cite{Adam:2015rna}. 
For non-spherical Skyrmions with $B>1$, we may define $\rho_B(r)$ and $\varepsilon(r)$ by averaging over the solid angles. However, (\ref{th_pres}) is hard to compute numerically in practice because for $B>1$ the radial derivatives vanish at some finite value of $r$, and only an exact cancellation of the zeros yields a finite result. { Actually, there is no fundamental reason why both densities should have critical points at the same value of the radial coordinate (although we have numerically checked that this seems to be the case, up to numerical accuracy), hence the `thermodynamical' definition of pressure (\ref{th_pres}) may not be well behaved in this sense when applied to Skyrmions.} Furthermore, away from this critical point,  the two definitions are seen to disagree as in  Fig.~\ref{fig:Pressures} (right)  in the case of $B=2$.

\section{Conclusions}

In conclusion, we have computed various definitions of radius for a variety of nuclei in the Skyrme model and studied their $B$-dependence. Their relations to the  D-term have been numerically confirmed to good precision.  These radii are associated with different components of the energy momentum tensor and have different physical interpretations. Together with the more familiar charge radius, they characterize the rich internal structure of the nucleon and nuclei. { Apart from characterizing their dependence with the baryon number for light and medium-sized Skyrmion solutions, we have unveiled a rather surprising property, namely, that the tensor radius (\ref{ten}) and the baryon charge radius (\ref{rhob1}) coincide for all the solutions we have considered up to a few percent. This points towards a non trivial relation between (the derivatives of) the associated form factors, namely, the $A(t)$ and $G_B(t)$, even when there is no apparent reason for the associated densities to be related, \emph{a priori}. }

We have also computed the  `pressure' distribution and found that $p(r)$ is negative in the core region of all the $B>1$ nuclei explored in this work. This is due  to the sign change of the $D(t)$ form factor at large $|t|$, and is  presumably related to the `hollowness' of the $B>1$  Skyrmions. Whether this is a general feature or an artifact of the model is unclear to us. It is therefore worthwhile to compute the present observables in more realistic approaches using the machinery of low energy nuclear physics. 

 A final remark concerning the semi-classical approximation adopted in this work is in order. It is well known that the  energy momentum tensor components get quantum corrections at first order in the $N_c^{-1}$ expansion due to the spin and isospin degrees of freedom. Thus, a natural extension of the work presented here would be to study how the different radii and their $B$-dependences get affected by such corrections. While we do not expect  large differences  from what  we have predicted here just by using the classical solutions, it is still interesting to study how the energy and pressure distributions differ between isobaric nuclei, as is the case with the charge density \cite{Danielewicz:2013upa}.

\begin{acknowledgements}
 A. G. M. C. acknowledges financial support  from the PID2021-123703NB-C21 grant funded by MCIN/ AEI/10.13039\\/501100011033/ and by ERDF, ``A way of making Europe''; and the Basque Government grant (IT-1628-22). M. H. G. is grateful to the Xunta de Galicia (Consellería de Cultura, Educación y Universidad) for the funding of his predoctoral activity through \emph{Programa de ayudas a la etapa predoctoral} 2021. Y.~H. is supported by the U.S. Department of Energy under Contract No. DE-SC0012704, and also by  Laboratory Directed Research and Development (LDRD) funds from Brookhaven Science Associates.
\end{acknowledgements}

\bibliography{Bibliography}

\end{document}